# THE PROBLEM OF THE "PREBIOTIC AND NEVER BORN PROTEINS"

Gerald E. Marsh

Argonne National Laboratory (Ret)
5433 East View Park
Chicago, IL 60615

E-mail:gemarsh@uchicago.edu

**Abstract.** It has been argued that the limited set of proteins used by life as we know it could not have arisen by the process of Darwinian selection from all possible proteins. This probabilistic argument has a number of implicit assumptions that may not be warranted. A variety of considerations are presented to show that the number of amino-acid sequences that need have been sampled during the evolution of proteins is far smaller than assumed by the argument.

The genetic code specifies 20 amino acids, and a typical protein might be made up of a sequence of 200 amino acids. The argument is often made that the probability is negligible that the relatively limited set of something like 20,000 proteins coded for in the human genome (or the far greater number found in the natural proteome) arose by Darwinian selection of random variations out of the $20^{200}$ possibilities (Chiarabelli and De Lucrezia 2007). As put by Chiarabelli and Lucrezia, "Nature could not have explored all possible amino acid combinations and, therefore, that many proteins with interesting new properties could have never been sampled by Nature. By and large, we are faced with the problem of how the 'few' extant proteins were produced and/or selected during prebiotic molecular evolution … Even knowing a useful method to produce proteins under prebiotic conditions the problem would not be solved. In fact, for example, synthesizing a random 50 mer chain using all 20 different amino acids it is theoretically possible to produce about $10^{65}$ different sequences and the probability to sample two identical chains is approximately equal to zero."

Notice that Chiarabelli and Lucrezia speak of proteins being "produced and/or selected during *prebiotic* molecular evolution". This means before the advent of self-reproducing organisms having the capability of both replication and metabolism. When and how the first proteins came into being is a critical issue for the origin of life and will be further discussed later in this essay.

The argument that the number of proteins found in nature could not have arisen by Darwinian selection of random variations brings to mind the Levinthal paradox (Levinthal 1969), which has to do with protein folding times. Levinthal's paradox results from assuming an unbiased random search, and can be resolved by introducing a small energy cost for locally incorrect bond configurations (Zwanzig, Szabo, and Bagchi 1992). In this way, the search is transformed into a biased search, which can dramatically reduce the number of configurations that need be sampled to arrive at a useful one.

From an evolutionary perspective, perhaps the most important proteins are the catalysts known as enzymes. What the probabilistic argument above tells us is that, if one assumes that the synthesis of each of the $20^{200}$ protein possibilities is equally probable, some of the earliest proteins capable of serving as enzymes could not have arisen from random selection from the set of all possible proteins. That is, they could not result from random variations in the sequence of amino acids followed by the natural selection of primitive organisms utilizing the resulting proteins. The resolution of this conundrum must lie with the origin of early self-reproducing systems. Like the Levinthal paradox, the evolution of a set of biologically useful proteins could not have depended on an unbiased selection from all possible proteins. The argument also tells us that during the early development of life, the set of possible proteins had to be sampled in a massively parallel manner.

The way cells currently produce proteins is the result of some three billion years of evolution. Even a cursory examination of the process shows that it is far too complex to have been used by early life: To begin with, amino acids go through the process of "activation", in which an amino-acid-specific aminoacyl-tRNA synthetase combines a given amino acid with adenosine monophosphate (AMP)—one of the nucleotides that make up RNA—and attaches the activated amino acid to one of some 35 different types of tRNA. These in turn are brought to the ribosome via the guanosine triphosphate (GTP)/guanosine diphosphate (GDP) cycle where mRNA (after the introns are spliced out) instructs the ribosome to produce a specific polypeptide chain. While a majority of proteins produced in this way may fold on their own, an important number use auxiliary folding proteins. This entire process could only have sprung full blown into existence by the intervention of some *deus ex machina*!

The principal difficulty with the evolutionary approach to understanding the problem with protein development is that while the synthesis of amino acids in the reducing atmosphere of the early earth was demonstrated in the laboratory as early as 1953 (Miller 1953; Miller and Orgel 1974), it is energetically unfavorable for amino acids to combine into polypeptides without the help of a catalyst. In modern cells, peptide bond formation is mediated by the energy in the amino acid-tRNA bond.

There is some recent work that could shed some light on this issue. In studying the origin and evolution of the ribosome (Harish and Caetano-Anollés 2012) it has been found that modern protein synthesis may have evolved from preexisting functions of primordial molecules. They found that universally conserved, functionally important components at the interface of the ribosomal small subunit and the large subunit are primordial.

It was the discoveries beginning in the early 1980s—that RNA could not only catalyze RNA replication but also direct peptide synthesis—that led to the idea of an "RNA World" (Orgel 2004) where life forms based on RNA existed before the ability to synthesize proteins from information encoded into DNA evolved. It was the recognition that the information contained in many eukaryotic genes was not contiguous and that introns had to be removed from the mRNA derived from these genes before the mRNA could be used for protein synthesis by the ribosome that led to these discoveries. Some RNAs were even found to be capable of self-splicing. Since the early work in this area, catalytic RNAs (known as ribozymes) have been identified that form amide bonds (known as peptide bonds in a biochemical context) between RNA and an amino acid or between two amino acids (Zhang and Cech 1998).

Both RNA-catalyzed aminoacyl-RNA synthesis and RNA-catalyzed amino acid activation have now been shown to be possible. Aminoacyl-tRNA synthetases catalyze two essential reactions: the activation of the carbonyl groups of amino acids by forming aminoacyl-adenylates designated as *aa-AMP* and the transfer of the aminoacyl group to a specific RNA. The first can be written as

$$aa + ATP \rightarrow aa\text{-}AMP + PP_i,$$

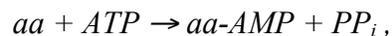

where *ATP* and *AMP* stand for adenosine triphosphate and adenosine monophosphate respectively. The anion $P_2O_7^{4-}$ is abbreviated as $PP_i$ and is formed by the hydrolysis of ATP into AMP. This reaction is followed by the transfer of the aminoacyl group to RNA

$$aa\text{-}AMP + RNA \rightarrow aa\text{-}RNA + AMP.$$

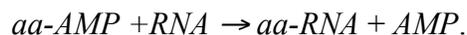

RNA-catalyzed aminocyl-RNA synthesis was first achieved by (Illangasekare, *et al*. 1995) and amino acid activation by (Kumar and Yarus 2001). The structural basis for specific tRNA aminoacylation by a small ribozyme has be given by (Hong, *et al*. 2008).

There is also at least one plausible prebiotic polymerization reaction that can produce peptides. It has been shown by (Leman, Orgel, and Reza Ghadiri 2004) that the volcanic gas carbonyl sulfide is capable of polymerizing amino acids to form peptides under reasonable conditions. Whether this reaction could be the basis for a complementary or alternative scenario to that of the RNA world has yet to be determined.

This brings us back to the issue of whether life first began in the form of creatures capable of rudimentary metabolism coupled with replication or, as put by Dyson in his delightful little book *Origins of Life* (Dyson 1985), whether "life began twice, with two separate kinds of creatures, one capable of metabolism without exact replication, the other kind capable of replication without metabolism". There is also the coevolution theory of the genetic code (Wong 2005) based on the postulate that prebiotic synthesis was an inadequate source of all twenty protein amino acids, and consequently some of them had to be derived from the coevolving pathways of amino acid biosynthesis.

Dyson introduced a "Toy Model" of molecular evolution along the lines of the Oparin picture of the origin of life (Oparin 1966) where proto-cells came first, followed by enzymes, and subsequently by the use of nucleic acids to store biological information. The model does not allow Darwinian selection. Its purpose was to demonstrate that a population of molecules within a proto-cell could, by random drift, achieve an organized state where active biochemical cycles might exist. Wong's theory, on the other hand, allows for a prebiotic evolution of peptide sequences as well as amino acid biosynthesis, and therefore at least partly makes use of Darwinian selection.

Without Darwinian selection, one cannot preferentially increase the number of proto-cells containing biologically useful molecules formed by random drift. Darwinian evolution can be introduced into Dyson's Toy Model by allowing the proto-cells to grow and reproduce by fission and dissolve in the absence of adequate nutrients. The precise extent and nature of prebiotic molecular evolution in the development of life is unknown, but a resolution of this question is not central to the issue at hand; being able to form proteins via the carbonyl sulfide or a similar route on the prebiotic Earth, or in an RNA World for that matter, does not in and of itself resolve the probabilistic argument given above.

Certain assumptions that may not at first be apparent are built into that argument. The first is that the evolution of life depends on selecting for the specific set of proteins that appears in life forms today. It seems very unlikely that the origin and evolution of life could be so narrowly constrained. The second is that each of the $20^{200}$ possible proteins is functionally unique. This is also unlikely to be the case. It is far more probable that many of these possible $20^{200}$ proteins can serve the same biological function. In the case of enzymes, it may have been possible for many structurally unrelated enzymes to catalyze a given reaction, albeit with possibly different reaction rates. This is certainly the case today. Such enzymes are known as "analogous" enzymes and represent independent paths in enzyme evolution (Galperin, et al. 1998; Gherardini et al. 2007). This contrasts with homologous enzymes, which derive from a common ancestor. It is also probable that many of the enzymes of the RNA World were far less efficient than their modern cognates.

Another assumption implicit in the argument that the protein space to be sampled is ~$20^{200}$ is that an enzyme can catalyze only a single reaction. The active site of an enzyme is generally only a very small part of the protein and is usually formed by different polypeptide chains. Even the configurations of enzymes often change in the presence of the substrate molecule with which it interacts. It is quite possible that many of the $20^{200}$ proteins, especially the larger ones, could have several active sites catalyzing different reactions. In fact, many enzymes are known to have a number of catalytic sites (Llewellyn and Spencer 2007). Taking account of these built in assumptions could dramatically reduce the set of possible proteins that need be sampled to produce an enzyme with a given activity.

There is another even more compelling argument that the set of proteins that was sampled during the early evolutionary process was far smaller than $20^{200}$. Nineteen of the twenty amino acid residues in the secondary structures of proteins show a relatively strong tendency to form either $\alpha$-helices, $\beta$-sheets or turns [See Berg, et al. (2002)]. That is, amino acids have different conformational propensities for forming these structures. Thus, from the secondary structural point of view, amino acids fall into three "fuzzy" equivalence classes. "Fuzzy" because, strictly speaking, an equivalence relation partitions a set into subsets where each member of the original set belongs to only one member of the partition. Here, amino acid residues only have a "tendency" to form $\alpha$-helices, $\beta$-sheets or turns.

In addition, it has been shown (Wei, et al. 2003) that the secondary structure of proteins depends not only on the composition but also the sequence of amino acids in the protein. Some sequence patterns produce $\alpha$-helical structures while others produce $\beta$-strands.

Most naturally occurring proteins are composed of between 50 and 2000 amino acids. If the size of early proteins was toward the lower end of this range, and if the tertiary structures upon which the protein's properties depend were strongly influenced by this division into three fuzzy amino acid equivalence classes, the set of proteins that would need to be sampled could be reduced. This reduction, in conjunction with the rules that specify the secondary structure of proteins, would dramatically reduce the sequence space that needs to be sampled. While the remaining number of proteins may still be large, it is one that could well be sampled in a reasonable time if the set of proteins is sampled in a parallel manner as discussed below.

Another consideration that could reduce the set of possible proteins is the recognition that more than one-third of all enzymes contain either bound metal ions or require the addition of such ions for catalytic activity. Since 1932, when carbonic anhydrase—a biologically important zinc containing enzyme—was discovered, hundreds of enzymes have been found to contain zinc, but only in the +2 state.

Zinc atoms in such enzymes are essentially always bound to four or more ligands. Carbonic anhydrase, which catalyzes carbon dioxide hydration, is particularly important in biological systems. The way this enzyme probably works is that the zinc creates a hydroxide ion from a water molecule by facilitating the release of a proton; the carbon dioxide substrate then binds to the enzyme's active site where it is positioned to react with the hydroxide ion; the hydroxide ion then converts the carbon dioxide molecule into a bicarbonate ion; and finally, the catalytic site is restored by the release of the bicarbonate ion and the binding of another water molecule.

A synthetic analog of this carbonic anhydrase mechanism, where a simple synthetic ligand binds zinc through four nitrogen atoms—rather than the three histidine nitrogen atoms in the enzyme—accelerates the hydration of carbon dioxide by more than a factor of 100 at a pH of 9.2 (Berg, Tymoczko, and Stryer 2002).

Zinc is not the only metal ion of interest. It has also been proposed (van der Gulik, et al. 2009) that specific short peptides three to eight amino acids long bound to one or more positively charged metal ions such as $Mg^{2+}$ could have served as catalysts during the period of very early life.

What this suggests is that early proto-enzymes may have contained metal complexes that were subsequently incorporated into evolving protein enzymes. If true, this too could have diminished the number of possible proteins that need be sampled—provided large subsets of the possible proteins were able to incorporate a given metal complex.

Thus, what the probability argument is really saying is that the possibility of life arising for a second time in exactly the same way it did is negligible. This may well be true, but many alternative possibilities may have existed for life to begin in the form of self-replicating systems. It is even possible that more than one type of self-replicating organism appeared and all were subject to variation, selection, and in the end, possible convergence.

The idea of an RNA World has its own difficulties. Two of the most important are obtaining, under pre-biotic conditions, nucleotides in sufficient quantity, and the lack of a known mechanism for replication of RNA molecules without the presence of a replicase. These challenges have not yet been fully resolved. However, once self-replicating RNA molecules capable of catalyzing polypeptide chains exist, it would be expected that different strands of RNA would produce different polypeptides. Collectively, these effectively sample the set of proteins in a massively parallel manner. Once the minimal necessary set of ribozymes were formed, even if their reaction rates were much lower than their modern cognates, RNA would serve to carry both "genetic" information and serve as a catalyst for the reactions needed for the first primitive self-replicating systems.

These primitive RNA World proto-cells would be subject to the usual Darwinian variation and selection process. There is, however, a strong incentive for an evolutionary transition from an RNA to a DNA world (Lazcano et al. 1988). This is due to the greater genetic stability of the double stranded helical structure of DNA compared to single stranded RNA, and the fact that deoxyribose is chemically less reactive than ribose.

In all modern organisms, the components of DNA are synthesized from those of RNA by ribonucleotide reductases. These convert the base and phosphate groups linked to a *ribose* sugar that form ribonucleotides to deoxyribonucleotides composed of a base and phosphates linked to *deoxyribose* sugar. The fact that ribonucleotide reductases take a variety of species dependent forms increases the probability that a primordial form of these reductases could have formed in the RNA World. Another possibility is that ribozymes were replaced by enzymes formed of amino acids before the evolutionary transition to storing information in DNA.

The transition to a DNA World is complicated by the fact that some mechanism for reading the information from the DNA would have had to evolve along with the transition to information storage in DNA.

One approach to resolving this problem might be along the following lines: The two strands that comprise DNA separate on heating, and reform on cooling. If RNA is present during the cooling period, DNA-RNA hybrids will form where sections of the DNA and the RNA have complimentary bases. Under natural selection, primitive self-reproducing proto-cells of the RNA World would have a preponderance of the type of RNA needed to support their metabolism and replication. If some of these RNAs were converted to DNAs by early forms of ribonucleotide reductases, a replication of RNAs using these DNAs as a template by temperature cycling may serve in place of the modern enzyme RNA polymerase. Once such DNA-based self-replicating systems existed, natural selection would rapidly end the RNA world.

It is interesting, and relevant to this discussion, that the Central Dogma has itself been questioned. Francis Crick, who coined the phrase in 1957, apparently views RNA editing as the most significant exception (Thieffry and Sarkar 1998). Such editing involves the modification of RNA sequences after transcription and is a common occurrence in eukaryotic cells. Although editing implies that the sequence of amino acids in the resulting protein is not entirely encoded in nucleic acids, there is still no known mechanism where the information for an amino acid sequence could flow from protein to nucleic acid.

RNA editing is common in mRNA transcribed from mitochondrial DNA. It was soon discovered, however, that the sequence information needed for RNA editing was supplied by small RNAs transcribed from a second component of the mitochondrial DNA. If this turns out always to be the case, RNA editing would involve the transfer of genetic information from one RNA to another (Seiwert 1996). This would resolve the editing challenge to the Central Dogma. Other challenges to the Central Dogma, such as the discovery of reverse transcriptase, the replication of prions, and epigenetic modulation of DNA, have been discussed elsewhere—for example, by (Morange 2008).

To summarize: If proteins on the average are composed of 200 amino acids, one possible way out of the dilemma of selecting the 20,000 or so proteins coded for in our genes (or the larger number found in the natural proteome) out of the $20^{200}$ possible is to reduce the set of possible proteins by recognizing that: it may have been possible for organisms that first populated the post-RNA World to use many different subsets of the $20^{200}$ possible proteins; when these proteins are enzymes, many more than one of them may catalyze any given reaction, although possibly with differing reaction rates. In addition, many proteins may have more than one active site and thus serve to catalyze more than one reaction. Also, of the $20^{200}$ possible, many proteins may be able to serve the same biological function. The set of proteins that need be sampled may also have been reduced if many of the early protein-based enzymes were able to incorporate the metal complexes that may have constituted earlier proto-enzymes. And, finally, if the tertiary structure of early proteins was strongly influenced by the division of amino acids into three fuzzy equivalence classes, the number of proteins that need be sampled could be closer to ~$3^{50}$ rather than $20^{200}$. The basic point is that the evolution of proteins could not, and need not have involved an unbiased random search of all possible proteins.

Once the transition to a DNA world is accomplished, the set of possible proteins would continue to be sampled by variation and selection of early DNA-based cells. These, as many single-celled organisms do today, would presumably share genes through horizontal gene transfer. Because of this mode of information exchange, the evolutionary process of variation and selection becomes a massively parallel rather than a serial process. This is what allows colonies of microorganisms to rapidly adapt to changing environmental conditions today.